\documentclass[prl,twocolumn,
superscriptaddress,
 amsmath,amssymb,
 aps,floatfix
]{revtex4-2}
\usepackage{gensymb}
\usepackage{graphicx}
\usepackage{dcolumn}
\usepackage{bm}
\usepackage{multirow}
\usepackage{textcomp, gensymb}
\usepackage
{}
\newcommand{\bfa}{BaFe$_2$As$_2$}
\newcommand{\bna}{BaNi$_2$As$_2$}

\newcommand{\tc}{$T_c$}
\newcommand{\ts}{$T_S$}

\newcommand{\kz}{$k_z$}
\newcommand{\kx}{$k_x$}
\newcommand{\ky}{$k_y$}
\begin{document}


\title{Spectral Evidence for Unidirectional Charge Density Wave in Detwinned BaNi$_2$As$_2$}

\author{Yucheng Guo}
\affiliation{%
 Department of Physics and Astronomy, Rice University, Houston, 77005 Texas, USA\\
}%

\author{Mason Klemm}
\affiliation{%
 Department of Physics and Astronomy, Rice University, Houston, 77005 Texas, USA\\
}%

\author{Ji Seop Oh}
\affiliation{%
 Department of Physics and Astronomy, Rice University, Houston, 77005 Texas, USA\\
}%
\affiliation{%
 Department of Physics, University of California, Berkeley, 94720 California, USA\\
}%

\author{Yaofeng Xie}
\affiliation{%
 Department of Physics and Astronomy, Rice University, Houston, 77005 Texas, USA\\
}%

\author{Bing-Hua Lei}
\affiliation{%
 Department of Physics and Astronomy, University of Missouri, Columbia, 65211 Missouri , USA\\
}%

\author{Sergey Gorovikov}
\affiliation{%
Canadian Light Source Inc., University of Saskatchewan, Saskatoon, SK S7N 2V3, Canada\\
}%

\author{Tor Pedersen}
\affiliation{%
Canadian Light Source Inc., University of Saskatchewan, Saskatoon, SK S7N 2V3, Canada\\
}%

\author{Matteo Michiardi}
\affiliation{%
Department of Physics and Astronomy, University of British Columbia, Vancouver, BC V6T 1Z1, Canada\\
}%
\affiliation{Quantum Matter Institute, University of British Columbia, Vancouver BC V6T 1Z4, Canada}
\affiliation{Max Planck Institute for Chemical Physics of Solids, N\"{o}thnitzer Straße 40, 01187 Dresden, Germany}

\author{Sergey Zhdanovich}
\affiliation{%
Department of Physics and Astronomy, University of British Columbia, Vancouver, BC V6T 1Z1, Canada\\
}%
\affiliation{Quantum Matter Institute, University of British Columbia, Vancouver BC V6T 1Z4, Canada}

\author{Andrea Damascelli} 
\affiliation{%
Department of Physics and Astronomy, University of British Columbia, Vancouver, BC V6T 1Z1, Canada\\
}%
\affiliation{Quantum Matter Institute, University of British Columbia, Vancouver BC V6T 1Z4, Canada}

\author{Jonathan Denlinger}
\affiliation{%
Lawrence Berkeley National Laboratory, Berkeley, 94720 California, USA
}%

\author{Makoto Hashimoto}
\affiliation{%
Stanford Synchrotron Radiation Lightsource, SLAC National Acelerator Laboratory, Menlo Park, 94025 California, USA
}%

\author{Donghui Lu}
\affiliation{%
Stanford Synchrotron Radiation Lightsource, SLAC National Acelerator Laboratory, Menlo Park, 94025 California, USA
}%

\author{Sung-Kwan Mo}
\affiliation{%
Lawrence Berkeley National Laboratory, Berkeley, 94720 California, USA
}%

\author{Rob G. Moore}
\affiliation{%
Materials Science and Technology Division,Oak Ridge National Laboratory, Oak Ridge, 37831 Tennessee, USA
}%

\author{Robert J. Birgeneau}
\affiliation{%
 Department of Physics, University of California, Berkeley, 94720 California, USA\\
}%

\author{David J. Singh}
\affiliation{%
 Department of Physics and Astronomy, University of Missouri, Columbia, 65211 Missouri , USA\\
}%
\affiliation{%
 Department of Chemistry, University of Missouri, Columbia, 65211 Missouri, USA\\
}%

\author{Pengcheng Dai}
\affiliation{%
 Department of Physics and Astronomy, Rice University, Houston, 77005 Texas, USA\\
}%

\author{Ming Yi}
\email{mingyi@rice.edu}
\affiliation{%
 Department of Physics and Astronomy, Rice University, Houston, 77005 Texas, USA\\
}%

\date{\today}%

\begin{abstract}
The emergence of unconventional superconductivity in proximity to intertwined electronic orders is especially relevant in the case of iron-based superconductors. Such order consists of an electronic nematic order and a spin density wave in these systems. 
BaNi$_2$As$_2$, like its well-known iron-based analog BaFe$_2$As$_2$, also hosts a symmetry-breaking structural transition that is coupled to a unidirectional charge density wave (CDW), providing a novel platform to study intertwined orders. Here, through a systematic angle-resolved photoemission spectroscopy study combined with a detwinning $B_1g$ uniaxial strain, we identify distinct spectral evidence of band evolution due to the structural transition as well as CDW-induced band folding. 
In contrast to the nematicity and spin density wave in BaFe$_2$As$_2$, the structural and CDW order parameters in BaNi$_2$As$_2$ are observed to be strongly coupled and do not separate in the presence of uniaxial strain. Our measurements point to a likely lattice origin of the CDW in BaNi$_2$As$_2$.

\end{abstract}

\maketitle


Quantum materials hosting unconventional superconductivity tend to develop complex phase diagrams where multiple electronic orders interact. 
In the Fe-based superconductors (FeSCs), the ubiquitous intertwined order takes form in a C$_4$-rotational symmetry-breaking nematic phase and a 
spin density wave (SDW) 
~\cite{Wang2015-eb,Wang2016-my,Dai2015-mg}. 
The nematic order manifests in a tetragonal to orthorhombic structural transition, identified as electronically driven by a divergent nematic susceptibility from elastoresistance measurements~\cite{Chu2012-dk}. Additionally, rotational symmetry-breaking is observed in the electronic, magnetic and optical properties~\cite{Yi2011-xr,Zhao2009-rg,Chuang2010-cf,Homes2020-rv,Dhital2012-vo,Tanatar2010UniaxialstrainDO,Baek2016-ji,Niedziela2011-pp}. 
Superconductivity emerges when these competing orders are suppressed by either doping or pressure, resulting in \tc\ as high as 40 K~\cite{Takahashi2008-yl,Mizuguchi2010-bs}.

\begin{figure*}
\includegraphics[width=0.99\textwidth]{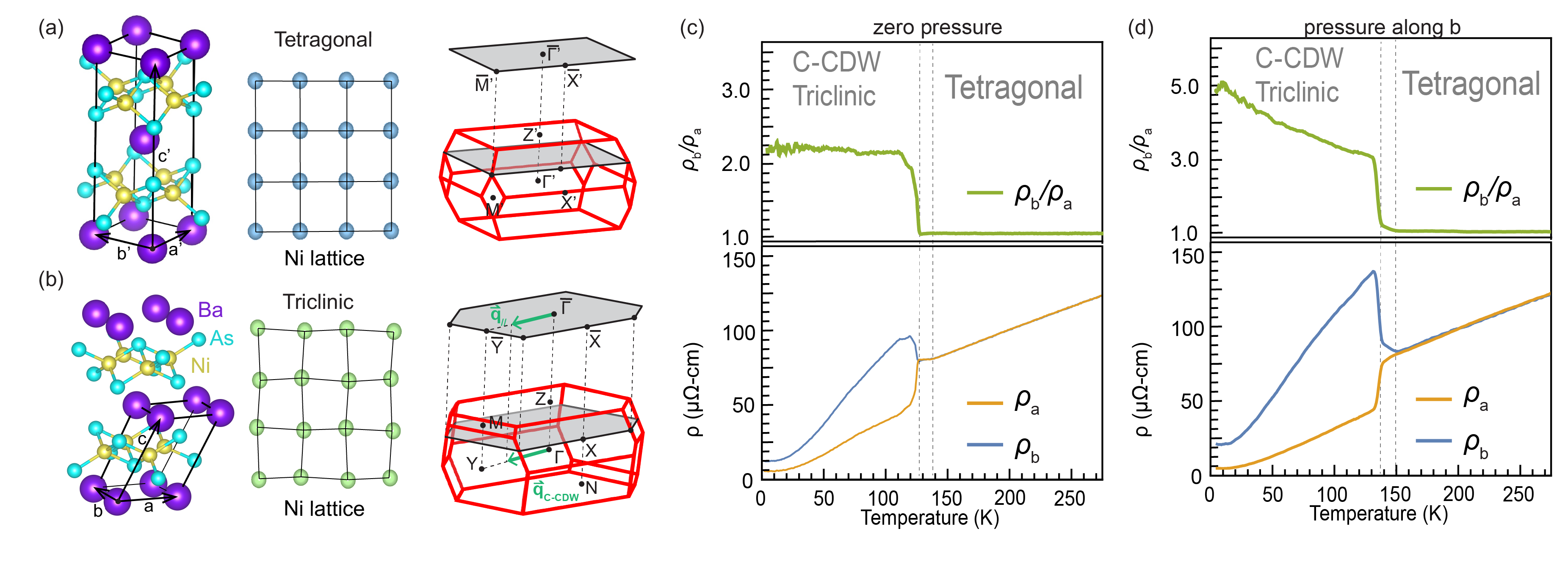}
\caption{\label{fig:Fig1} (a)-(b) Crystal structure and BZ for the tetragonal and triclinic phases. The red (gray) BZ represents the 3D (2D projected) BZ. The green arrows in (b) denote the q-vectors for C-CDW and its in-plane component. (c)-(d) Temperature-dependent in-plane resistivity $\rho_a$ (yellow), $\rho_b$(blue) and  $\rho_b$/$\rho_a$(green) with and without uniaxial compressive strain along b.}
\end{figure*}

BaNi$_2$As$_2$--a nickel-based analog of the well-studied FeSC \bfa--shares the same high temperature tetragonal phase~\cite{Sefat2009-xn}. Distinct from \bfa, \bna~is nonmagnetic and superconducts with a \tc~of 0.6K~\cite{osti_962242}. Instead of the coupled nematicity and SDW found in \bfa, \bna~exhibits a symmetry-lowering structural transition into a triclinic phase at \ts~=136 K accompanied by unconventional charge density waves (CDWs). The progression from the tetragonal state to the triclinic state is still under debate. From x-ray diffraction, Lee. \emph{et al.} discovered that the CDW first appears in the tetragonal state as an incommensurate order (IC-CDW), then transits to a different unidirectional IC-CDW at \ts~and then locks into a commensurate order (C-CDW) below \ts~\cite{Lee2019-uz,Lee2021-aj}. The crystal structure in this picture changes from tetragonal directly to triclinic at \ts~in a first order transition. Recent dilatometry work suggests that the crystal structure explicitly breaks C$_4$ rotational symmetry 
in the form of a orthorhombic phase in a second order transition before additional symmetry-lowering into the triclinic phase at \ts~in a first-order fashion~\cite{Merz2021-vn}. Substitution of either Co on the Ni site or Sr on the Ba site can completely suppress the triclinic phase and the associated CDW orders, reaching a maximum \tc~of 3.5 K. Elastoresistance measurements show that Ba$_{1-x}$Sr$_{x}$Ni$_2$As$_2$ exhibits a diverging nematic susceptibility in the $B_{1g}$ channel, which may be the cause for the enhancement of \tc~near optimal doping~\cite{Eckberg2020-ud}. 
A Ginzburg–Landau analysis suggests that the divergence of nematic susceptibility in Ba$_{1-x}$Sr$_{x}$Ni$_2$As$_2$ could be driven by either lattice or electronic degrees of freedom,
in contrast to BaFe$_2$As$_2$, where the divergence in the ${B}_2g$ channel is electronically-driven. 
\bna~therefore offers a rich platform to investigate intertwined orders, where the electronic structure could provide important insights into the nature of nematicity, CDWs, and their connection to unconventional superconductivity. 
Here, using angle-resolved photoemission spectroscopy (ARPES) under uniaxial strain, we probe the electronic structure evolution of BaNi$_2$As$_2$ across the phase transitions. We find no evidence of Fermi surface nesting in the tetragonal state, unidirectional band folding consistent with the reported C-CDW q-vector in the C-CDW/triclinic state, and rotational symmetry-breaking 
that onsets abruptly at the triclinic transition. Furthermore, the extracted temperature evolution of the spectral features identifies a distinct order parameter for the C-CDW from the triclinic structural transition. However, the two order parameters are observed to be strongly coupled even in the presence of uniaxial strain, distinct from the case of \bfa~where strain lifts the nematic band splitting to onset well above that of the SDW phase. Our results taken together suggest a strongly lattice-driven intertwined order in parent \bna~and much weaker nematic fluctuations compared to that of the FeSC \bfa.

\begin{figure}
\includegraphics[width=0.45\textwidth]{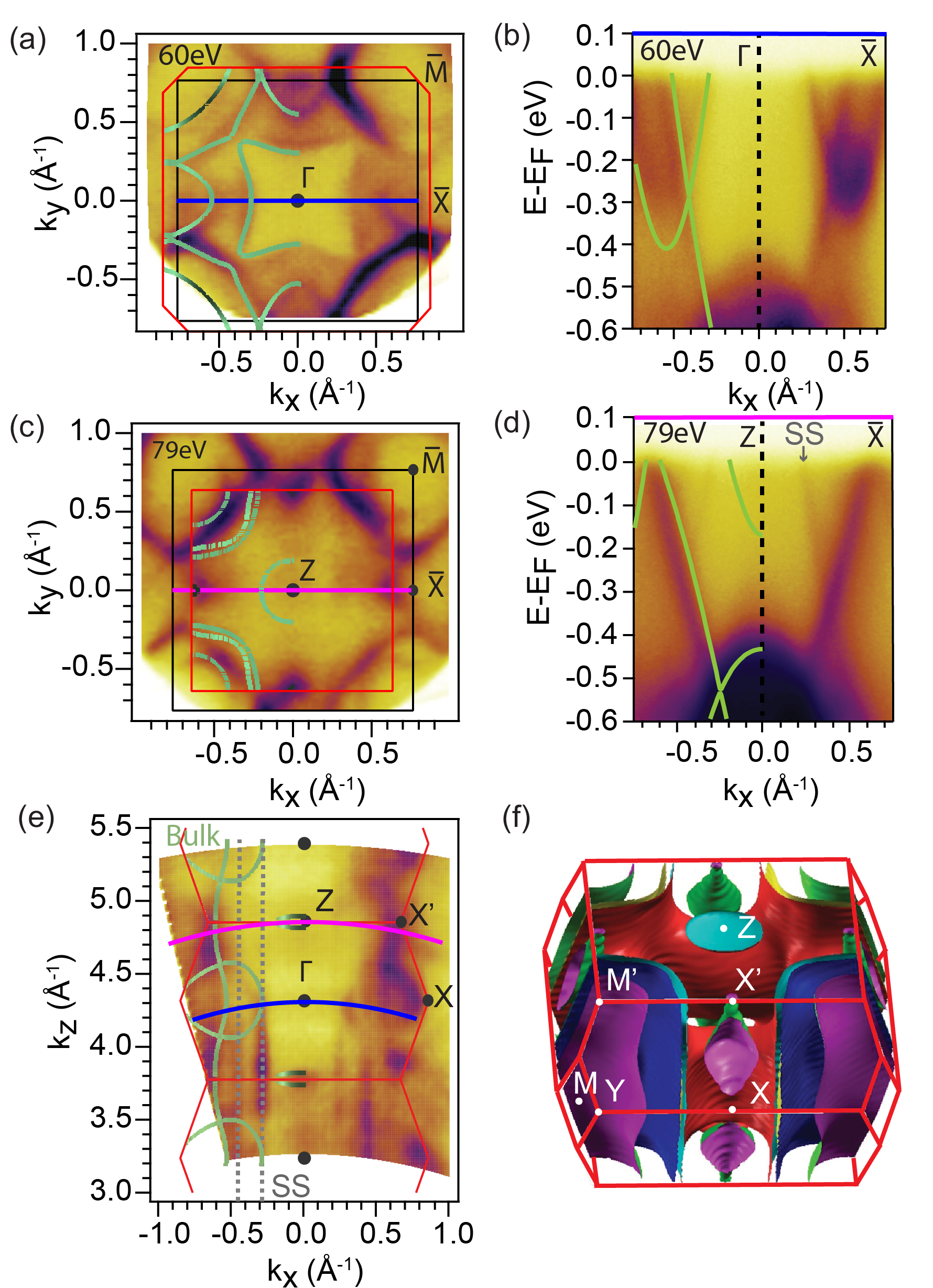}
\caption{\label{fig:Fig2}Electronic structure in the tetragonal state (170 K). (a) Fermi surface (FS) measured with 60 eV photons (\kz~=0). (b) Band dispersions measured along $\Gamma$-X as marked in (a). (c)-(d) Same as (a)-(b) but measured with 79 eV photons (\kz~=$\pi$). (e) Photon energy-dependent measurement in the k$_x$-k$_z$ plane, with an inner potential of 15 eV. (f) DFT calculation of the 3D FSs. In addition, DFT calculated band dispersions and FSs are overlaid on all data in green solid lines. Grey arrow and dashed lines mark the surface states. All data taken with circular right (CR) polarization.
}
\end{figure}

High-quality single crystals of BaNi$_2$As$_2$ were synthesized using the self-flux method~\cite{Sefat2009-xn}. Resistivity measurements were carried out in a Quantum Design Physical Properties Measurement System using a clamp to supply an in-plane, uniaxial pressure.~\cite{Chu2010-vb} ARPES measurements were performed at the QMSC beamline at the Canadian Light Source, beamlines 4.0.3 and 10.0.1 of the Advanced Light Source, and beamline 5-2 of the Stanford Synchrotron Radiation Lightsource with electron analyzers at an energy and angular resolution of 12 meV and 0.1$\degree$, respectively. The samples were cleaved \emph{in-situ} with base pressures below 5×10$^{-11}$ Torr. The polarization used was circular left (CL) unless otherwise noted. To avoid the domain mixing effect, we used a mechanical clamp to detwin the BaNi$_2$As$_2$ crystals along the [010] direction, with a typical pressure of 5-10 MPa~\cite{Fisher2011-qn}.

At room temperature, tetragonal BaNi$_2$As$_2$ belongs to the space group I4/mmm (a=4.142\AA, c=11.65\AA) (Fig.~\ref{fig:Fig1}(a)). It undergoes a first-order phase transition to the triclinic structure (a=4.21\AA, b=3.99\AA, c=6.31\AA, $\alpha$=105.2\degree, $\beta$=108.6\degree, $\gamma$=89.3\degree, space group P{$\bar1$}) at T$_S$ = 136K (Fig.~\ref{fig:Fig1}(b)). The corresponding 3D Brillouin Zones (BZ) of the tetragonal and triclinic phases are shown accordingly, where the rotational C$_4$ symmetry is broken for the triclinic phase. The C-CDW q-vector in the triclinic state is denoted by q$_{C-CDW}$, which does not lie in the projected BZ plane while its in-plane component (q$_{//}$) is very close to one third of the projected tetragonal BZ (See SM). We carried out temperature dependent in-plane electrical resistivity measurements along with two orthogonal directions with and without uniaxial pressure. In contrast to \bfa~\cite{Chu2010-vb}, even in the absence of a detwinning stress, we observe resistivity anisotropy ($\rho_b$- $\rho_a$) below T$_S$, signaling a strong structural distortion and associated unequal domain populations. In the presence of stress, the resistivity anisotropy is enhanced, demonstrating that the applied uniaxial stress redistributes the domain populations. In addition, we observe a kink above \ts~in the zero-stress sample, which is likely the reported IC-CDW transition~\cite{Lee2019-uz}. In the measurement under stress, the kink feature is replaced by the onset of the resistivity anisotropy, which can be interpreted as either a detwinning effect of the crystal orthorhombic domains~\cite{Yi2011-xr}, or strain-induced rotational symmetry-breaking~\cite{Song2019-wu}. It is interesting to point out that, in contrast to \bfa, the uniaxial pressure does not broaden the structural transition~\cite{Chu2010-vb,Tam2019-gw}.

To visualize the electronic structure of BaNi$_2$As$_2$, we present ARPES measurements taken in the tetragonal phase.
The Fermi surface maps (FSMs) and band dispersions along high symmetry directions correspond to the k$_z$=0 (Fig.~\ref{fig:Fig2}(a)-(b)) and k$_z$ =$\pi$/2 (Fig.~\ref{fig:Fig2}(c)-(d)) planes are taken with 60 eV and 79 eV photons, respectively (Fig.~\ref{fig:Fig2}(e)). The C$_4$ symmetry of the tetragonal state is observed, and there is overall good agreement with Density Functional Theory (DFT) calculations, consistent with previous ARPES reports on \bna~\cite{Zhou2011-gl,Pavlov2021-rn}. 
The electronic structure exhibits strong k$_z$ dispersion. The most intense Fermi surfaces are large pockets around the M points.
Besides dispersive bulk states, there exists surface states which are not dispersive along k$_z$ and have no correspondence in the bulk band calculations (denoted by SS, see SM for more details).

\begin{figure}
\includegraphics[width=0.45\textwidth]{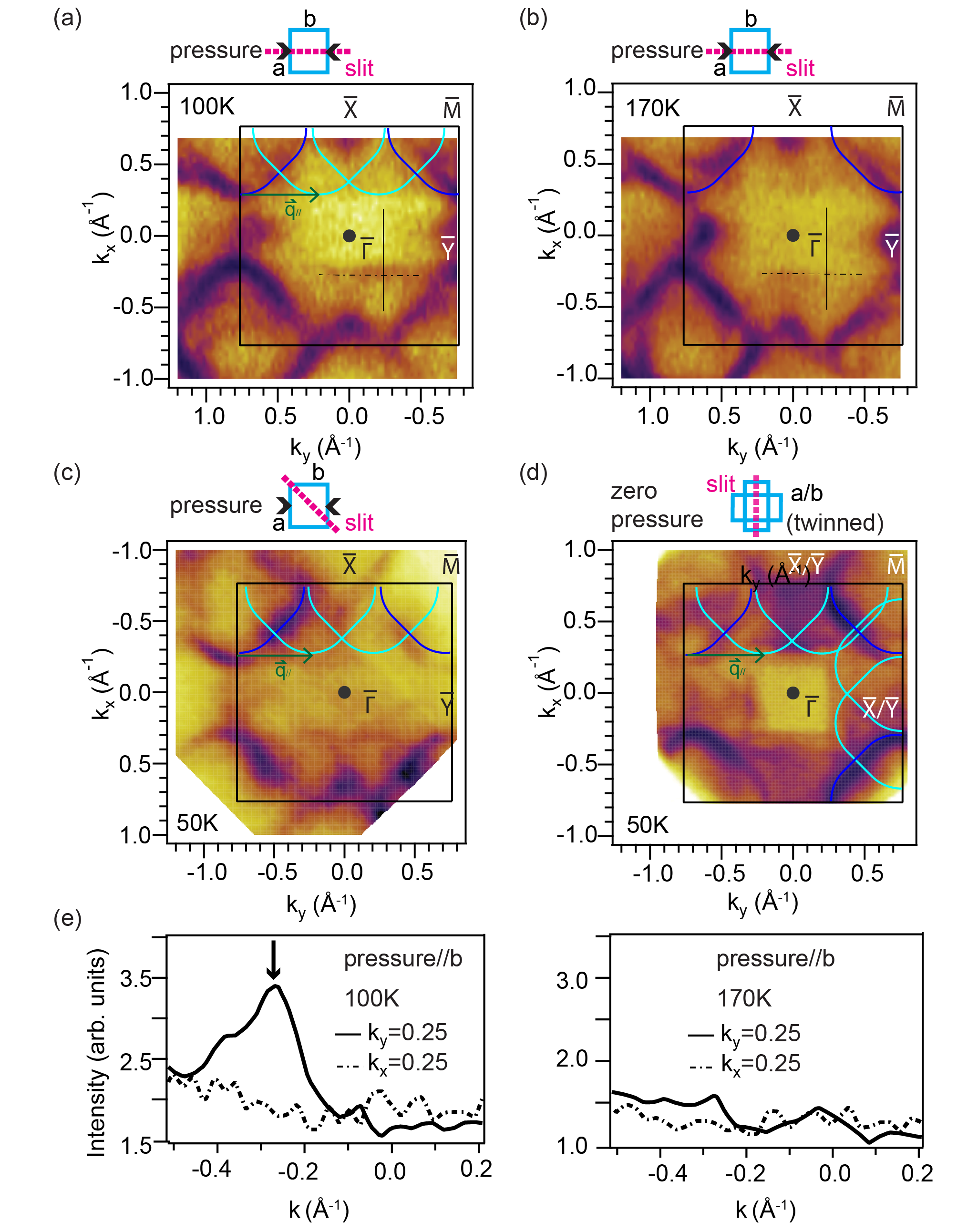}
\caption{\label{fig:Fig3}FSs of detwinned \bna. (a) FS taken at 79 eV in the triclinic phase (100K) with horizontal compressive uniaxial pressure. (b) Same as (a) but taken in the tetragonal phase (170K). (c) Same as (a) but taken at 50K with a 45{$\degree$} analyzer slit direction. (d) FS taken with CR 60eV photons at 50K without uniaxial pressure.Blue lines mark the Fermi pockets around $\bar{M}$, while cyan lines indicate bands folded from the $\bar{M}$ points due to the C-CDW. (e) MDCs along k${_y}$=0.25\AA$^{-1}$ (solid line) and $k_x$=0.25\AA$^{-1}$ (dashed line) for strained sample at 100K and 170K respectively, as marked in (a)-(b). The black arrow denotes the folded feature.}
\end{figure}

Having identified the electronic structure in the tetragonal state, we now examine the low temperature triclinic phase. To probe the intrinsic single-domain electronic structure, we apply uniaxial compressive strain along the crystal axis [010], which aligns all the domains along the shorter $b$-axis. The main features of the low temperature FSM of the strained crystal (Fig.~\ref{fig:Fig3}(a)) resemble their tetragonal counterparts except for the appearance of new bands marked in cyan, which can be understood as folded copies of the M pocket that only appear along the $b$-direction, hence breaking the rotational C$_4$ symmetry. This anisotropy can be demonstrated by a comparison of the momentum distribution curve (MDC) taken at equivalent momenta along the \kx~and \ky~directions, where only a peak is seen across the folded feature (Fig.~\ref{fig:Fig3}(e)).
Such band folding is a signature of translational symmetry breaking. We note that the folding vector is approximately $\frac{1}{3}$, which is consistent with the in-plane projection of the $q_{C-CDW}$ observed by x-ray diffraction~\cite{Lee2019-uz,Pavlov2021-rn}. Therefore, this band folding is a signature of the unidirectional CDW, which was not observed in previous ARPES studies of \bna~\cite{Zhou2011-gl}.

To exclude rotational symmetry-breaking effect due to extrinsic photoemission matrix elements, we measured the strained sample in a geometry where the $a$ and $b$ directions are symmetric with respect to the analyzer slit (Fig.~\ref{fig:Fig3}). In this geometry, the photoemission matrix elements are equivalent along the $k_x$ and $k_y$ directions therefore any observed difference must be intrinsic to the band structure. We note that the folded bands still only appear along the strained direction, reflecting a true C$_2$ symmetry. In addition, above the CDW ordering temperature (Fig.~\ref{fig:Fig3}(b)), the folded bands in the strained sample disappear, restoring the C$_4$ symmetry, as also confirmed by the disappearance of the peak in the MDC (Fig.~\ref{fig:Fig1}(e)).
For comparison, the FSM of an unstrained twinned crystal (Fig.~\ref{fig:Fig3}(d)) shows folded bands in both directions, consistent with the understanding of unidirectional CDW folding under mixed domains.
The comparison between the FSMs of the strained and unstrained crystals clearly establishes that uniaxial strain is effective at detwinning the crystal and important for resolving the observed CDW band folding. We also note that besides the bulk bands described here, we observed a set of surface states (SS) that strongly reflects a broken C$_4$ symmetry observed in strained crystals, demonstrating the effectiveness of the applied strain (See SM). These states have no correspondence in the bulk DFT calculated bands and do not disperse with \kz, consistent with a surface origin. 

\begin{figure*}
\includegraphics[width=0.99\textwidth]{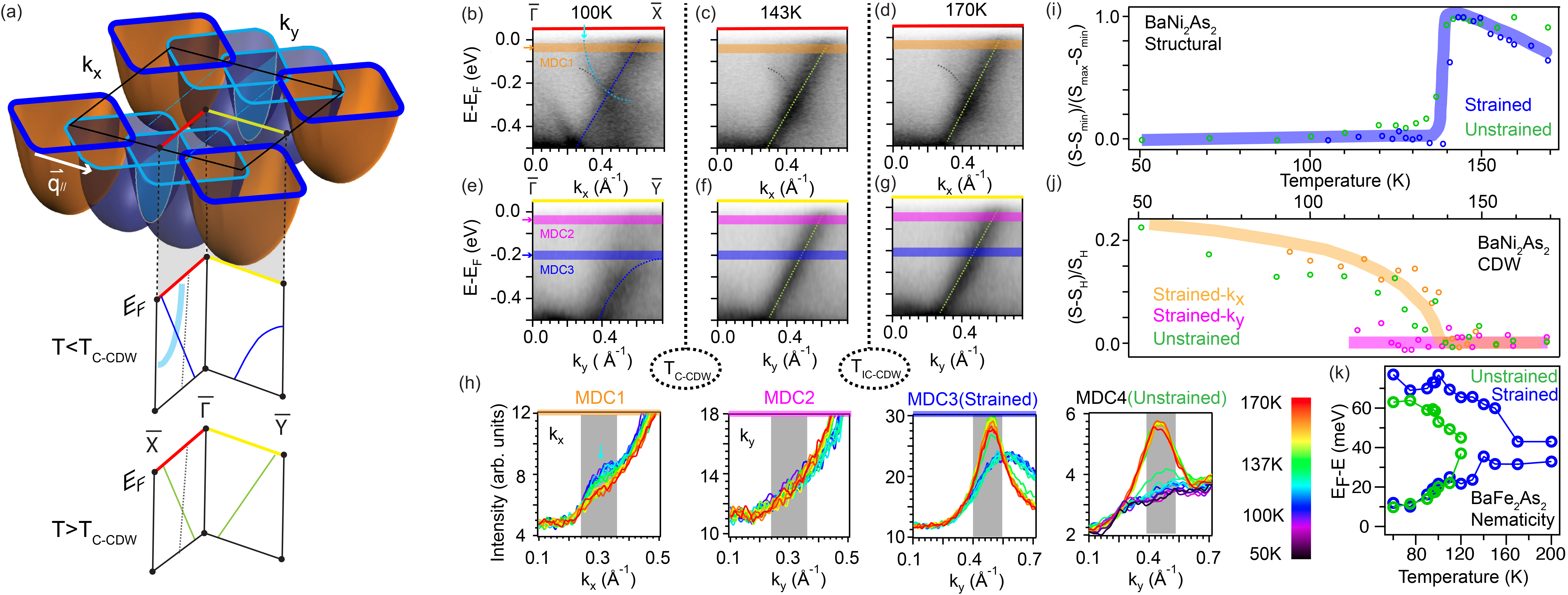}
\caption{\label{fig:Fig4}Temperature evolution through the CDW transitions on detwinned BaNi$_2$As$_2$. (a) Summary of the band evolution across T$_{C-CDW}$. (b)-(d) Temperature dependence of band dispersions along $\bar{\Gamma}$-$\bar{X}$ for (b) T $<$ T$_{C-CDW}$, (c) T$_{C-CDW}$ $<$ T $<$ T$_{IC-CDW}$, and (d) T $>$ T$_{IC-CDW}$. (e)-(g) Same as (b)-(d) but taken along $\bar{\Gamma}$-$\bar{Y}$. Dashed line are guides to the eyes. (h) Temperature dependence of MDCs as marked in (b)-(g). The locations of MDCs are as marked.
(i) Integrated spectral weight as a function of temperature within the gray window for MDC3 and MDC4. Each curve is normalized to the maximum spectral weight. (j) Integrated spectral weight within the gray window in MDC1 and MDC2 and the equivalent MDC taken on the unstained crystal as a function of temperature. Each curve is normalized by calculating (S-S$_H$)/S$_H$, where S$_H$ is the average spectral weight between 145K-170K. Solid lines are guides to the eyes. (k) Temperature evolution of d$_{xz}$ and d$_{yz}$ band positions in BaFe$_2$As$_2$ with and without uniaxial strain (reproduced from~\cite{Yi2011-xr}).
}
\end{figure*}

Next, we examine in detail the evolution of the rotational and translational symmetry breaking in the electronic structure.
Specifically, we traced the dispersions along the the $\bar{\Gamma}$-$\bar{X}$ and $\bar{\Gamma}$-$\bar{Y}$ directions measured on detwinned crystals as a function of temperature (Fig.~\ref{fig:Fig4}). While dispersions along the two orthogonal directions are identical in the tetragonal phase (170 K), bands are strongly modified in the low temperature phase (100 K), with band crossing only along $\bar{\Gamma}$-$\bar{X}$ but not along $\bar{\Gamma}$-$\bar{Y}$ (Fig.~\ref{fig:Fig4}(b)-(g). To understand the observed dispersions, we illustrate with a schematic in Fig.~\ref{fig:Fig4}(a). In the normal state above T$_{IC-CDW}$, the Fermi surface consists of large pockets around the $\bar{M}$ points. Two types of transitions occur that modify the bands. First, the structural transition from tetragonal to triclinic is reflected in a broken rotational symmetry, where the blue dispersion bends down along $\bar{\Gamma}$-$\bar{Y}$. Second, the unidirectional CDW folds the M pockets along \ky, resulting in a folded parabolic band (cyan) that only appears along $\bar{\Gamma}$-$\bar{X}$. 

We then use the spectral signatures of these two types of symmetry breaking to trace out their order parameters. First, we trace the spectral signature associated with the rotational symmetry breaking in the form of the band shift along $\bar{\Gamma}$-$\bar{Y}$. As captured by MDC3 in Fig.~\ref{fig:Fig4}(e),(h), an abrupt change occurs at \ts~in a strongly first-order fashion. This is also reflected in the apparent jump in the temperature-dependent spectral weight integrated across the gray momentum range (Fig.~\ref{fig:Fig4}(h)-(i)). Second, we trace the band folding by comparing the MDC across the cyan band for $\bar{\Gamma}$-$\bar{X}$ and $\bar{\Gamma}$-$\bar{Y}$ (Fig.~\ref{fig:Fig4}(h)). The presence of the folded band is seen in a hump in the MDC along $\bar{\Gamma}$-$\bar{X}$ (MDC1) but not $\bar{\Gamma}$-$\bar{Y}$ (MDC2). The spectral weight of the hump gradually decreases as the temperature is raised, confirmed by the integrated spectral weight in the marked momentum range plotted against temperature, which mimics the behavior of a second order phase transition (Fig.~\ref{fig:Fig4}(j)). This is in contrast to the abrupt jump of the rotational symmetry-breaking in Fig.~\ref{fig:Fig4}(i), suggesting that the CDW and triclinic structural transition have distinct order parameters.

Comparing the two extracted order parameters, we note that both appear to onset simultaneously in a single phase transition within our experimental uncertainty, although the CDW band folding order parameter evolves much more smoothly than that of the band shift. This contrast is also apparent in the collapsed MDC curves in Fig.~\ref{fig:Fig4}(h). Due to the abrupt onset, we ascribe the band shift transition to the first-order structural transition into the triclinic phase. While a small kink is observed above \ts~in the resistivity indicating the onset of IC-CDW and potential orthorhombic transition and nematic order, we do not observe any apparent change in the band dispersions above \ts~(see SM). This does not preclude the existence of a nematic phase above \ts, but rather indicates that the energy scale of the symmetry-breaking in the electronic structure associated with IC-CDW and nematic order is much smaller than that observed for the nematic phase in \bfa.

Finally, we study the effect of uniaxial strain on the order parameters extracted from the spectral signatures. The spectral weights acquired from unstrained crystals by following the same procedures described above are plotted in (Fig.~\ref{fig:Fig4}(i)-(j)), showing very similar behavior as those obtained on strained samples. In particular, we observe no elevation in the onset temperature of either order parameter in the warming up measurement.
This is again in contrast to \bfa~(Fig.~\ref{fig:Fig4}(k)), where the uniaxial strain clearly elevates the onset temperature of the observed orbital anisotropy and separates it from the SDW ordering temperature~\cite{Song2013-ht}. The results taken together suggest that the translational symmetry-breaking of the C-CDW and the structural transition into the triclinic phase are strongly coupled in \bna, and that nematic fluctuations are much weaker than that in \bfa. The C-CDW, from a lack of Fermi surface nesting conditions, is likely dominated by lattice. The recent DFT results show that two distinct structures compete due to complex As bonding patterns and drive distortions of the Ni layers which possibly explain the unconventional CDW behavior of the ground state~\cite{Lei2022}. 

In the broader context of FeSCs, the coupling of the nematic and magnetic order varies as well. In \bfa, the structural transition is second-order, followed by a first-order magnetic transition. In the structural homolog SrFe$_2$As$_2$, the structural and magnetic transitions are strongly first-order and occur simultaneously. Uniaxial strain also does little in harvesting the nematic fluctuations above \ts ~\cite{Tam2019-gw}. The case of \bna~as we demonstrate here, appears to be similar in spirit to that of SrFe$_2$As$_2$, where very weak rotational symmetry-breaking exists above the strongly first-order structural transition. However, with substitution of P on the As site or Sr on the Ba site, nematic fluctuations appear to grow both in the size of the anisotropic thermal expansion above \ts~\cite{Merz2021-vn}, as well as a diverging nematic susceptibility, which have been proposed to be responsible for the enhancement of \tc~\cite{Eckberg2020-ud}. \bna~therefore offers a rich platform analogous to the magnetic FeSCs where intertwined order from the charge-nematic sector interacts with superconductivity.


\section{acknowledgments}
ARPES experiments were performed at the Advanced Light Source and the Stanford Synchrotron Radiation Lightsource, which are both operated by the Office of Basic Energy Sciences, U.S. DOE. Part of the research described in this work was also performed at the Canadian Light Source, a national research facility of the University of Saskatchewan, which is supported by Canada Foundation for Innovation (CFI), the Natural Sciences and Engineering Research Council of Canada (NSERC), the National Research Council (NRC), the Canadian Institutes of Health Research (CIHR), the Government of Saskatchewan, and the University of Saskatchewan. The ARPES work at Rice University was supported by the Robert A. Welch Foundation Grant No. C-2024 and the Gordon and Betty Moore Foundation's EPiQS Initiative through grant No. GBMF9470. The materials synthesis efforts at Rice are supported by the US Department of Energy (DOE), Basic Energy Sciences (BES), under Contract No. DE-SC0012311 and the Robert A. Welch Foundation, Grant No. C-1839 (P.D.). Theory work at the University of Missouri was supported by the Department of Energy, BES, Award DE-SC0019114. Work at University of California, Berkeley and Lawrence Berkeley National Laboratory was funded by the U.S. Department of Energy, Office of Science, Office of Basic Energy Sciences, Materials Sciences and Engineering Division under Contract No. DE-AC02-05-CH11231 (Quantum Materials Program KC2202). The work at Oak Ridge National Laboratory was supported by the U.S. Department of Energy, Office of Science, National Quantum Information Science Research Centers, Quantum Science Center. This research was undertaken thanks, in part, to funding from the Max Planck-UBC-UTokyo Center for Quantum Materials and the Canada First Research Excellence Fund, Quantum Materials and Future Technologies Program.

\bibliography{BNA}

\end{document}



\title{Spectral Evidence for Unidirectional Charge Density Wave in Detwinned BaNi$_2$As$_2$: Supplemental Material}

\author{Yucheng Guo}
\affiliation{%
 Department of Physics and Astronomy, Rice University, Houston, 77005 Texas, USA\\
}%

\author{Mason Klemm}
\affiliation{%
 Department of Physics and Astronomy, Rice University, Houston, 77005 Texas, USA\\
}%

\author{Ji Seop Oh}
\affiliation{%
 Department of Physics and Astronomy, Rice University, Houston, 77005 Texas, USA\\
}%
\affiliation{%
 Department of Physics, University of California, Berkeley, 94720 California, USA\\
}%

\author{Yaofeng Xie}
\affiliation{%
 Department of Physics and Astronomy, Rice University, Houston, 77005 Texas, USA\\
}%

\author{Bing-Hua Lei}
\affiliation{%
 Department of Physics and Astronomy, University of Missouri, Columbia, 65211 Missouri , USA\\
}%

\author{Sergey Gorovikov}
\affiliation{%
Canadian Light Source Inc., University of Saskatchewan, Saskatoon, SK S7N 2V3, Canada\\
}%

\author{Tor Pedersen}
\affiliation{%
Canadian Light Source Inc., University of Saskatchewan, Saskatoon, SK S7N 2V3, Canada\\
}%

\author{Matteo Michiardi}
\affiliation{%
Department of Physics and Astronomy, University of British Columbia, Vancouver, BC V6T 1Z1, Canada\\
}%
\affiliation{Quantum Matter Institute, University of British Columbia, Vancouver BC V6T 1Z4, Canada}
\affiliation{Max Planck Institute for Chemical Physics of Solids, N\"{o}thnitzer Straße 40, 01187 Dresden, Germany}

\author{Sergey Zhdanovich}
\affiliation{
Department of Physics and Astronomy, University of British Columbia, Vancouver, BC V6T 1Z1, Canada\\
}%
\affiliation{Quantum Matter Institute, University of British Columbia, Vancouver BC V6T 1Z4, Canada}

\author{Andrea Damascelli} 
\affiliation{
Department of Physics and Astronomy, University of British Columbia, Vancouver, BC V6T 1Z1, Canada\\
}%
\affiliation{Quantum Matter Institute, University of British Columbia, Vancouver BC V6T 1Z4, Canada}

\author{Jonathan Denlinger}
\affiliation{%
Lawrence Berkeley National Laboratory, Berkeley, 94720 California, USA
}%

\author{Makoto Hashimoto}
\affiliation{%
Stanford Synchrotron Radiation Lightsource, SLAC National Acelerator Laboratory, Menlo Park, 94025 California, USA
}%

\author{Donghui Lu}
\affiliation{%
Stanford Synchrotron Radiation Lightsource, SLAC National Acelerator Laboratory, Menlo Park, 94025 California, USA
}%

\author{Sung-Kwan Mo}
\affiliation{%
Lawrence Berkeley National Laboratory, Berkeley, 94720 California, USA
}%

\author{Rob G. Moore}
\affiliation{%
Materials Science and Technology Division,Oak Ridge National Laboratory, Oak Ridge, 37831 Tennessee, USA
}%

\author{Robert J. Birgeneau}
\affiliation{%
 Department of Physics, University of California, Berkeley, 94720 California, USA\\
}%

\author{David J. Singh}
\affiliation{%
 Department of Physics and Astronomy, University of Missouri, Columbia, 65211 Missouri , USA\\
}%
\affiliation{%
 Department of Chemistry, University of Missouri, Columbia, 65211 Missouri, USA\\
}%

\author{Pengcheng Dai}
\affiliation{%
 Department of Physics and Astronomy, Rice University, Houston, 77005 Texas, USA\\
}%

\author{Ming Yi}
\email{mingyi@rice.edu}
\affiliation{%
 Department of Physics and Astronomy, Rice University, Houston, 77005 Texas, USA\\
}%

\date{\today}%

\maketitle

\section{CDW q vectors}

\begin{figure}
\includegraphics[width=0.3\textwidth]{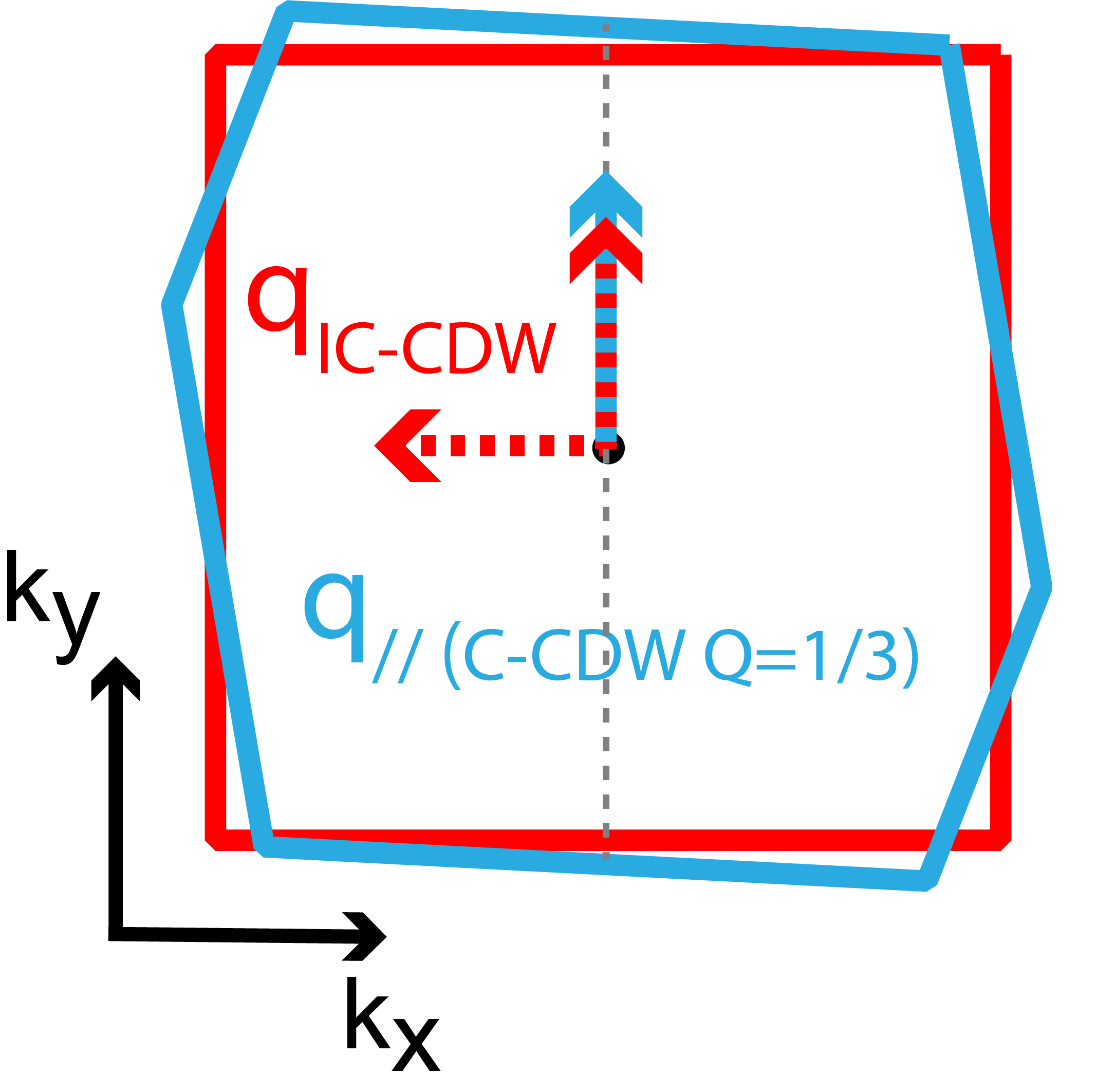}
\caption{\label{fig:FigS1} In-plane components of CDW q vectors for the triclinic and tetragonal phases. The red BZ represents the 2D projected BZ for the tetragonal phase while the blue BZ is the 2D projected BZ for the triclinic phase.}
\end{figure}
The band folding in the triclinic state \bna~is not straightforward. The folding vector is along the b$_2$ direction q=(0,1/3,0) in the momentum space.  This q vector has both in-plane and out-of-plane components. For the bands with large \kz~ dependence, the folded bands on the Fermi surface should not be simple replica of the original bands as they are from different \kz. However, there is one exception for \bna~as the electron pockets round the M points don't exhibit large \kz~dependence as has been shown in Fig. 2. Therefore, the folded bands of the big M pockets resemble the original band and appear as the replica in Fig. 3. In Fig. 3, we plotted the projected BZ for the tetragonal state instead of the triclinic state for simplicity. The folding q vector is close to one third of the BZ‘s side length (0.51~$\AA^{-1}$) which is approximately equal to 1/3 of the b$_2$ vector (0.525~$\AA^{-1}$) of the triclinic state (Fig.~\ref{fig:FigS1}).

\section{Strain effect on the surface state}
\begin{figure}
\includegraphics[width=0.35\textwidth]{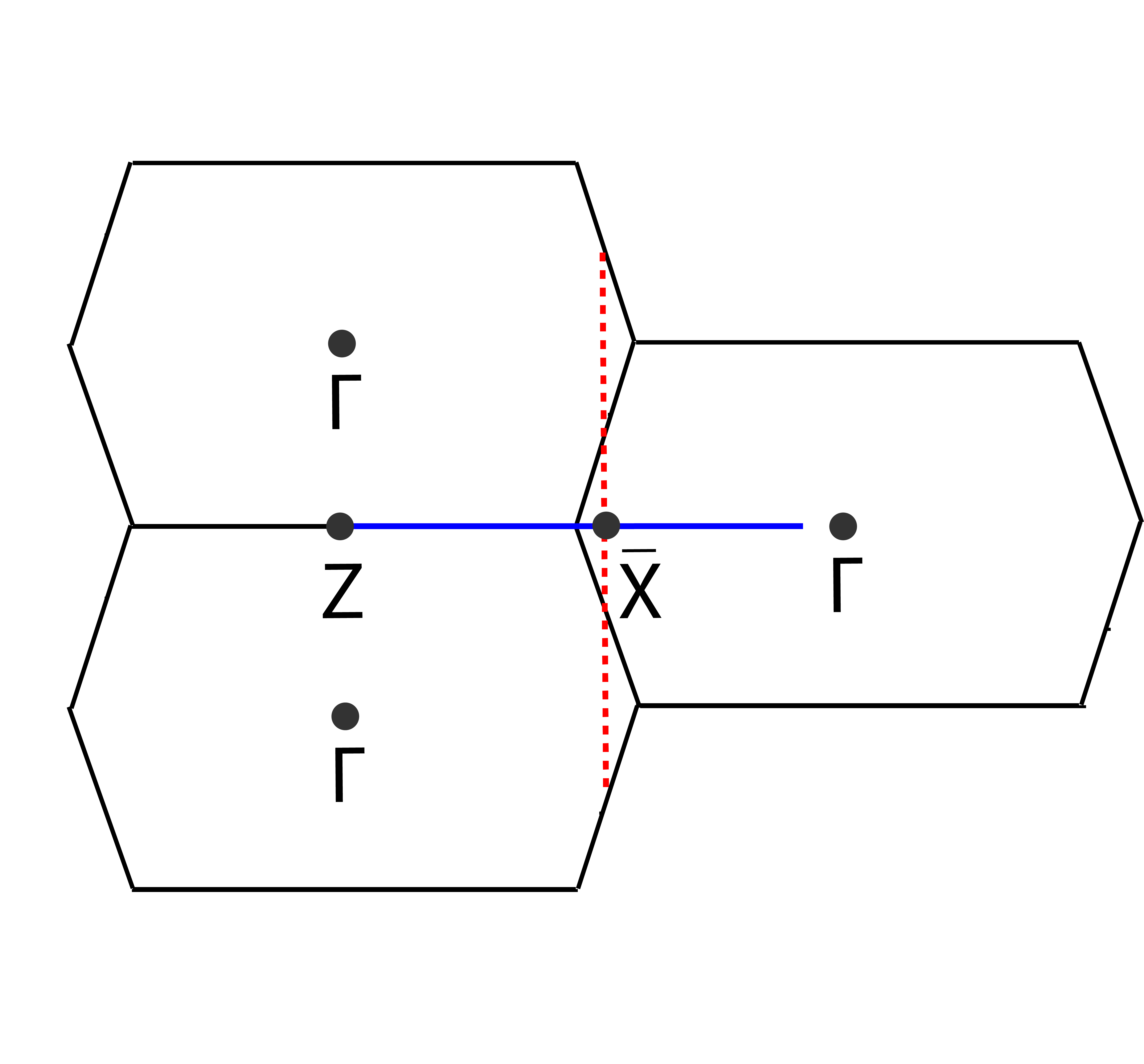}
\caption{\label{fig:FigS2} Illustration of unsymmetrical bulk band k$_z$ across $\bar{X}$ in band dispersion taken with photon energy corresponding to k$_z$=$\pi/2$.
}
\end{figure}

\begin{figure*}
\includegraphics[width=0.95\textwidth]{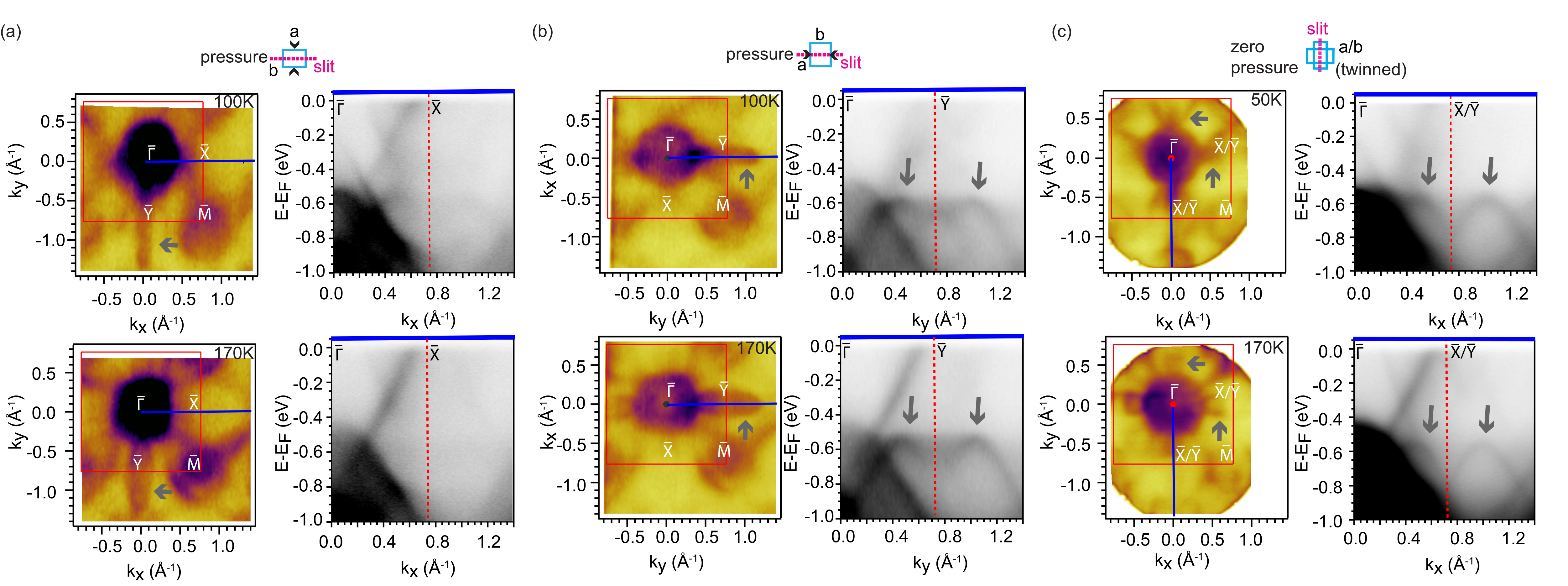}
\caption{\label{fig:FigS3} Strain induced anisotropic surface states of BaNi$_2$As$_2$. (a) Constant energy contour at E$_B$=0.6 eV and band dispersions measured along the $\bar{\Gamma}$-$\bar{X}$ direction at 100 K and 170 K with vertical compressive uniaxial pressure. (b) Same as (a) but taken with horizontal compressive uniaxial pressure and the band dispersions are measured along the $\bar{\Gamma}$-$\bar{Y}$ direction. (c) Same as (a) but taken without uniaxial pressure at 50 K and 170 K respectively. The gray arrows mark the surface state symmetrical across $\bar{X}$.
}
\end{figure*}

As described in the main text, the bulk bands always resume C$_4$ rotational symmetry when the system warms up into the tetragonal state shown in Fig. 3 and Fig. 4. However, the surface states in \bna~have a different behavior. There are three ways to tell if a band has a surface origin: (a) The band is not dispersive in the photon energy scan (Fig. 2), (b) The high temperature spectrum is not captured by the bulk band dispersion, (c) The band is symmetric across the BZ boundary, which also indicates that the band is not dispersive along \kz~since for body-centered tetragonal structure, extending out from the first BZ enters the next BZ at a different \kz~(Fig.~\ref{fig:FigS2}). In the presence of uniaxial strain, we find that the constant energy contour at E$_B$=0.6 eV always appears to be anisotropic for all temperature measured up to 170 K, which is well into the tetragonal state (Fig.~\ref{fig:FigS3}(a)(b)). The high symmetry cuts show that the anisotropy in the constant energy contours arise from bands that are symmetrical about the BZ boundary. These bands only appear along $\bar{\Gamma}$-$\bar{Y}$ but not along $\bar{\Gamma}$-$\bar{X}$. Together with the fact that these bands are not captured by the DFT calculation in Fig. 2, we conclude they have surface origins and very sensitive to the uniaxial strain. In a strained sample, these surface states only exist along the strain direction. For unstrained sample, the surface bands hold C$_4$ symmetry in both states (Fig.~\ref{fig:FigS3}(c)). Besides, the domain mixing effect of bulk bands described in the main text can be clearly seen in the comparison between Fig.~\ref{fig:FigS3}(a)-(c).

\begin{figure*}
\includegraphics[width=0.95\textwidth]{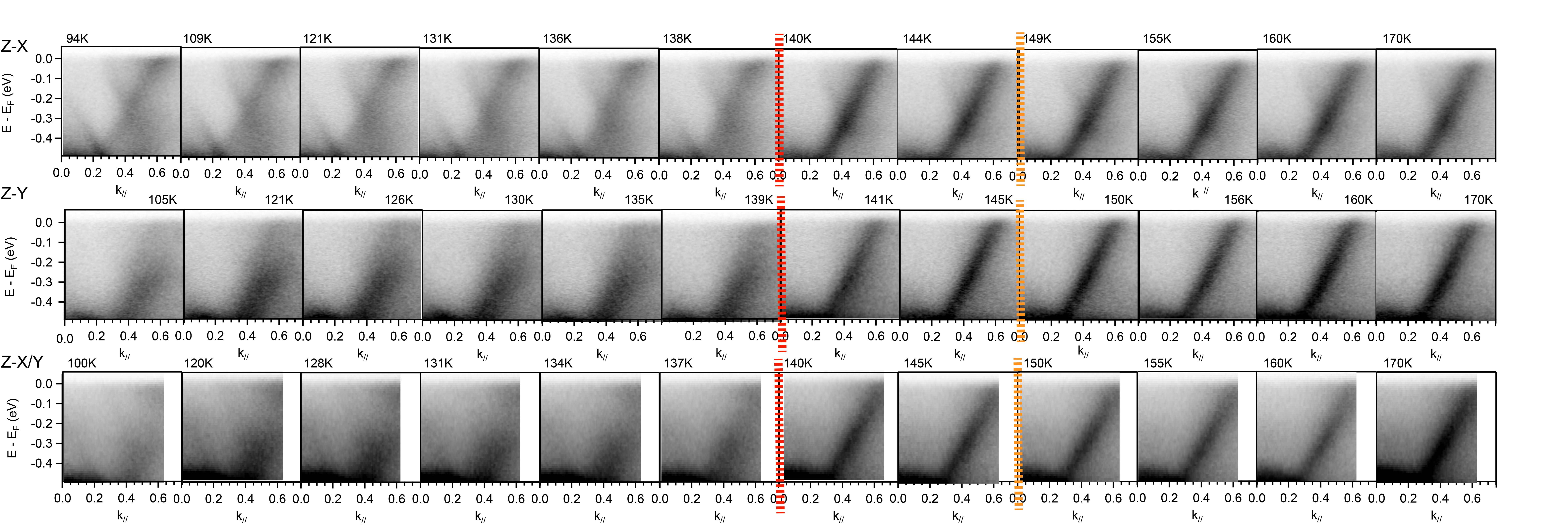}
\caption{\label{fig:FigS4} Temperature evolution of the high symmetry cuts taken on strained and unstrained samples. first row: $Z-X$ (strained); second row: $Z-Y$(strained); third row: $Z-X/Y$ (unstrained). T$_S$ is denoted by the red dashed line while T$_{ICDW}$ is denoted by the orange dashed line.
}
\end{figure*}

\section{temperature evolution of high symmetry cuts}

Raw spectra from which we extracted the order parameter temperature evolution showed in Fig. 4 are presented in Fig.~\ref{fig:FigS4}. Two strongly coupled order parameters were extracted. To note, unlike the spectra shown in Fig.~\ref{fig:FigS3}(c), the domain mixing effect for the unstrained spectra shown in Fig.~\ref{fig:FigS4} does not have an equal distribution of the two domains, similar to the observation of the resistivitiy anisotropy of the unstrained sample (Fig. 1(c)) and is therefore not a direct sum of the spectra of $\bar{\Gamma}$-$\bar{X}$ and $\bar{\Gamma}$-$\bar{Y}$. There is more domain population for $\bar{\Gamma}$-$\bar{Y}$ than $\bar{\Gamma}$-$\bar{X}$, causing the $\bar{\Gamma}$-$\bar{Y}$ spectra to have a higher intensity in the mixed signal. Meanwhile, the uneven distribution does not affect the MDC analysis and the spectral weight extraction, as the peaks (Fig. 4) are well separated in momentum space in the analysis.

\bibliography{BNA}